\documentclass[12pt]{iopart}
\usepackage{iopams}
\usepackage{graphicx}
\usepackage{cite}
\eqnobysec
\begin{document}

\title[Intermittent random walks for optimal search
strategy]{Intermittent random walks for an
optimal search strategy: One-dimensional case}

\author{G.Oshanin$^1$, H.S.Wio$^2$, K.Lindenberg$^3$, and
S.F.Burlatsky$^4$}

\address{$^1$ Physique Th\'eorique de
la Mati\`ere Condens\'ee (UMR 7600), Universit\'e Pierre et Marie
Curie - Paris 6, 4 place Jussieu, 75252 Paris France and Department
of Inhomogeneous Condensed Matter Theory, Max-Planck-Institute f\"ur
Metallforschung, Heisenbergstrasse 3, D-70569 Stuttgart, Germany}

\address{$^2$ Instituto de Fisica de Cantabria, Avda. Los Castros
s/n, E-39005 Santander, Spain}

\address{$^3$ Department of Chemistry and
Biochemistry 0340 and Institute for Nonlinear Science,
University of California, San Diego, La Jolla, CA
92093-0340, USA}

\address{$^4$ United Technologies Research Center, United
Technologies Corp., 411 Silver Lane, 129-21 East Hartford, CT 06108,
USA}

\ead{oshanin@lptl.jussieu.fr, wio@ifca.unican.es,
kl@hypatia.ucsd.edu, burlatsf@utrc.utc.com}

\begin{abstract}
We study the search kinetics of an immobile target by a concentration
of randomly moving searchers. The object of the study is to optimize the
probability of detection within the constraints of our model.
The target is hidden on a one-dimensional lattice in the sense that searchers
have no \textit{a priori} information about where it is,
and may detect it only upon encounter. The searchers perform
random walks in discrete time $n=0,1,2, \ldots, N$, where $N$ is
the maximal time the search process is allowed to
run. With probability $\alpha$ the searchers step on a
nearest-neighbour, and with probability $(1-\alpha)$ they leave the
lattice and stay off until they land back on the lattice at a
fixed distance $L$ away from the departure point. The random walk is thus
\textit{intermittent}.  We calculate
the probability $P_N$ that the target remains undetected up
to the maximal search time $N$, and seek to minimize this probability.
We find that $P_N$ is
a non-monotonic function of $\alpha$, and show that there is an optimal
choice $\alpha_{opt}(N)$ of $\alpha$ well within the intermittent regime,
$0 < \alpha_{opt}(N) < 1$, whereby $P_N$ can be orders of
magnitude smaller compared to the ``pure" random walk cases
$\alpha =0$ and $\alpha = 1$.
\end{abstract}

\submitto{\JPC}

\maketitle

\section{Introduction}
\label{introduction}

Everybody searches for something: predators look for
prey, prey forage, human beings look for better
places to work or for lost keys or for partners, spies search for
hidden secrets, and knights desperately seek the
Holy Grail, whatever it may be.

The search for a desired target may be long and uncertain; targets
may be sparse, hidden or difficult to detect even when found, or
they may have their own life-time and vanish before they are
detected. That is why efficient search strategies appropriate to
specific types of situations are highly desirable. The question of
optimal search strategies has motivated a great deal of work within
the last
years~\cite{ste,sto,bell,frost,yossi,vis1,leva,bart,rap,boyer,ben,mike}.

While earlier work has focused on systematic searches in organized
human activities such as rescue
operations~\cite{ste,sto,bell,frost}, more recent analyses have been
devoted to random
strategies~\cite{yossi,vis1,leva,bart,rap,boyer,ben}. In particular,
it has been realized that in the case of the so-called
non-destructive search, when the target reappears after some time at
the same location, and when the searcher always remains within the
system, L\'evy flights with randomly reoriented ballistic
trajectories provide an optimal search
efficiency~\cite{yossi,vis1,leva,bart,rap}. On the other hand,
following the observation of trajectories of foraging animals such
as lizards or fish or birds characterized by two distinct types of
motion -- fast relocation stages non receptive to the target and
relatively slow \textit{reactive} phases when the target may be
detected~\cite{obr,kra} -- another type of random search, an
intermittent search, has been proposed. Analytical modeling of such
random intermittent search strategies has been put forth
in~\cite{ben} (see also ~\cite{mike}).

In this paper we discuss the search of a single target hidden on a
one-dimensional regular lattice by a concentration of searchers
performing \textit{intermittent} random walks. In particular, at
each tick of the clock each of the searchers chooses between two
possibilities. One is to jump to a nearest-neighbouring lattice
site, and the other is to leave the lattice and fly with a given
velocity until it lands back on the lattice at a fixed distance $L$
from the departure site. The term ``hidden" means that the searchers
can detect the target only upon landing directly on the target site
at the end of a flight, or via a one-step (nearest-neighbour) walk
onto that site. Intermittency has been invoked in other models
(e.g., see the models proposed in~\cite{ben,mike}) and has been
observed in a number of ecological and other contexts (see, e.g.,
~\cite{obr,kra}), but all the other features of our model are, to
the best of our knowledge, new.

Contrary to previous work, which focused on the analysis and
optimization of the first passage time to the target from a given
location, or on the number of targets
encountered~\cite{yossi,vis1,leva,bart,rap,boyer,ben}, here we study
the behaviour of a different property, namely, the probability that
a single target remains \textit{undetected} up to the maximal search
time $N$. We can not, of course, make this random search process
certain; that is, we can not guarantee that the target is found (or
not found) with unit probability in a finite time, but what we are
able to show is that the non-detection probability is a
non-monotonic function of the parameter $\alpha$ which determines
whether a searcher takes a nearest-neighbour step (probability
$\alpha$) or a long flight [probability $(1-\alpha)$], and that it
has a sharp minimum at some value of $\alpha$. Consequently, we show
that the search efficiency can be dramatically enhanced, by orders
of magnitude, by choosing an appropriate value of $\alpha$.

The outline of this paper is as follows. In section~\ref{model}
we describe our model of intermittent random walks,  present
definitions, and delineate our main results.
In section~\ref{nondetection} we derive a general formula for the
non-detection probability. Section~\ref{intermittent} is devoted to
the analysis of
different properties of the intermittent random walk. In
section~\ref{particular}
we focus on a particular exactly solvable case, namely, a Lindenberg-Shuler
intermittent walk with steps of unit length and flights of two
lattice spacings. For this example we discuss the specific features
of the optimization procedure. Next, in section~\ref{optimal} we
consider the
general case of intermittent random walks with steps of unit length
and flights of fixed length $L$, and define an optimal search strategy.
Finally, in section~\ref{conclusions} we conclude with a brief
summary of our results and an outlook of future work.

\section{The model, basic notation and results}
\label{model}

Consider a one-dimensional regular lattice of unit spacing
containing $M+1$ sites $s$. This is the substrate. At one of the lattice sites,
say at the origin $s = 0$, we hide an immobile target, and only we
know that it is there. Then, we randomly place $K$ searchers
in the lattice under the constraint that none of them is placed
at the site with the hidden target, that is, we distribute them randomly
over the remaining $M$ sites. The searchers have no knowledge
of the location of the target.

\begin{figure}[!htb]
\begin{center}
\includegraphics[width = 0.8\textwidth,height=0.2\textwidth]{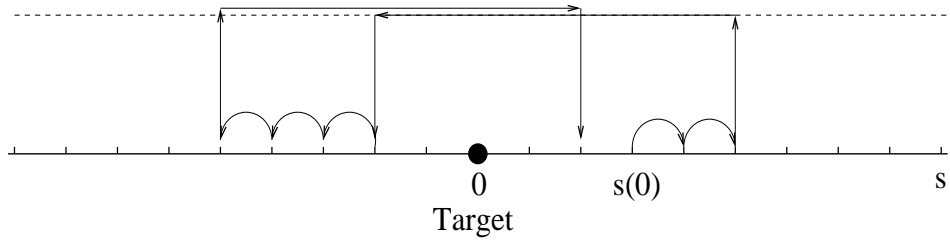}
\caption{\label{Fig1} A sketch of a $7$-step trajectory of an
intermittent random walk, starting at $s(0)$, with
nearest-neighbor steps (probability $\alpha$) and off-lattice
flights [probability $(1-\alpha)$] over a distance $L$ (here, $L = 7$).
The arrows indicate the step direction. Note that this trajectory does
not find the target.}
\end{center}
\end{figure}

Next, we let the searchers move according to the following rule
(see figure~\ref{Fig1}): At each tick of the clock, $n =1,2,3, \ldots, N$
($N$ is the maximal time the search process may run), each
searcher selects randomly between two possibilities.
With probability $\alpha$, the searcher jumps to one of its nearest
neighbouring sites, with equal probabilities to the right or to the
left. With probability $(1 - \alpha)$, the searcher leaves the lattice
and flies off-lattice with a given velocity $V$ until it lands at a
site an fixed integer distance $L$ away from the departure
site. The direction of the flight is chosen at random, with equal
probabilities to right or left. The time the searcher spends
off-lattice during
a flight is $T = L/V$, a model parameter that can have integer
values $T = 1,2, \ldots, L$. Note that this value defines the
velocity $V$.

As stated earlier, the term ``hidden" means that a searcher can not
perceive the target when it is off-lattice and may detect it only
when it lands on the target site.  The searcher may land there
at the end of a flight or by a one-step jump.  Once the searcher lands
on the target site, one can define an
elementary detection probability $q$. The target may
immediately be recognized when first encountered by a searcher
($q \equiv 1$, perfect detection),
or detection may take place with a given probability $0 < q < 1$.
In the latter case, when
any of the searchers  arrives at the site occupied by the target,
there is always a possibility that the target remains undetected
(imperfect detection). Here we focus on the perfect detection case.
Some aspects of the search kinetics in the imperfect
detection case for the one-dimensional system will be discussed
briefly in section~\ref{conclusions}. An extension of the model to
$d$-dimensional spaces and some other subtle questions not pursued here
will be discussed elsewhere~\cite{we2}.

Our first goal is to determine the probability $P_N$ that the
target remains undetected up to time $N$, the maximal time
the search process is allowed to run, which merely depends on our
patience or on experimental constraints. We show in the next section
that in the perfect detection
limit the probability $P_N$ that no searcher has reached the target up
to time $N$ obeys the standard formula (see, e.g.~\cite{t1,t2} for
more details)
\begin{equation}
P_N = \exp\left(- \rho S_N\right),
\label{survival}
\end{equation}
where $S_N$ is the expected value of the \textit{number of distinct
sites} visited on the substrate by any searcher in an
$N$-step walk (see, e.g.,~\cite{hughes}).  \emph{This is a crucial
equation}, since we will arrive at results for $P_N$ via
calculations of $S_N$.

We will calculate $S_N$ explicitly for the intermittent random walk
and determine its asymptotic behaviour analytically in
the large-$N$ limit. Using these results, we set out to show that
there always exists
a value $\alpha = \alpha_{opt}(N)$, $0 < \alpha_{opt}(N)<1$,
which maximizes the expected number of distinct
substrate sites visited by an intermittent random walk up to
time $N$. Consequently, it minimizes the
probability $P_N$ in equation~(\ref{survival}).  We
demonstrate that by choosing this optimal value of $\alpha$ the
non-detection probability $P_N$ in equation~(\ref{survival}) can be made
orders of magnitude smaller than those of the ``pure" cases $\alpha =1$
(nearest-neighbor steps only) or $\alpha = 0$ (flights only),
for which asymptotically
\begin{equation}
\label{clas2}  P_N(\alpha =1) = \exp\left[- \rho \left(\frac{8
N}{\pi}\right)^{1/2} + {\cal
O}\left(\frac{1}{\sqrt{N}}\right)\right].
\end{equation}
and
\begin{equation}
\label{clas1}  P_N(\alpha =0) = \exp\left[- \rho \left(\frac{8
N}{\pi T}\right)^{1/2} + {\cal
O}\left(\frac{1}{\sqrt{N}}\right)\right],
\end{equation}
The result~(\ref{clas2}) is well-known and reflects the fact that a
random walk in low dimensional systems ($d = 1$ and $d = 2$)
compactly explores the space; that is, $S_N$ grows
\textit{sublinearly} with time $N$, and each site ever visited by
such a walk is most probably visited many times. Hence, searching a
target in a low dimensional system with the help of a standard random
walk is not very efficient, since the walker
wastes a lot of time revisiting sites that do not contain the
target. The result~(\ref{clas2}) is the same with the replacement $N\to
N/T$ (i.e., it is the same as that of a nearest neighbor random walk for
a shorter time).  This further decreases the probability of detection
because it does not explore the local regions at all before making a
jump to a site that has probably been visited many times before.  It
is thus not an efficient search strategy either.
We find that although the sublinear growth of $S_N$ is unavoidable in
any one dimensional random walk, and, even more specifically, the
asymptotic dependence $S_N\propto N^{1/2}$, for fixed $L$ it is
possible to optimize the search
strategy via an intermittent walk with a particular choice of $\alpha$
that depends on the length of the walk (calculated explicitly later).  We
obtain the asymptotic result
\begin{equation}
\label{}
\fl
P_N(\alpha = \alpha_{opt}(N))
\sim \exp\left[- \rho
\left(L V\right)^{1/2} \left(\frac{8 N}{\pi}\right)^{1/2} + \rho
\gamma(L) \left(\frac{8 N}{\pi}\right)^{1/6} + o\left(N^{1/6}\right)
\right],
\end{equation}
where $\gamma(L)$ is an increasing function of the flight distance
$L$. Note that the leading term in the exponent
contains a pre-factor $(L V)^{1/2}$, $V=1,2,3, \ldots,L$,
which can make it considerably larger than the corresponding terms in
equations~(\ref{clas1}) and (\ref{clas2}) and thus may result in a
substantial increase of the detection probability.

\section{Non-detection probability}
\label{nondetection}

In this section we define the probability $P_N$ that the target is
not detected up to time $N$, and express it using well-known
properties of random motion that have been extensively discussed in the
literature (see, e.g.,~\cite{t2,hughes} and references therein).
In particular, we begin by introducing the notation
$s(n)$ to denote the random process that describes a searcher's
trajectory, and $P(s|s(0);n)$ to denote the probability that $s(n)$, which
is initially at site $s(0)$, is at site $s$ at time $n$.
We find it useful to present the derivation because,
contrary to most studies, we are interested in defining
the number of distinct sites visited by $s(n)$ only on the substrate,
which is a subset of all available space. Indeed, most existing
work concerns random walks that never leave the lattice so that
the equality $\sum_s P(s|s(0);n) \equiv 1$
holds at all times.  In our case this probability is not conserved,
$\sum_s P(s|s(0);n) < 1$, since the searchers spend some portion of
their time off-lattice. We present this analysis in the most
general case, assuming only that $s(n)$ is a homogeneous process.

Suppose there are altogether $K$ searchers labeled $k=1,2,\ldots,K$,
and let $s_k(n)$  denote a given $n$-step trajectory of
the $k$-th searcher. Then, in the perfect detection case  the
indicator function $\Psi\Big(\{s_k(N)\}\Big)$ of the event that the
target has not been detected up to time $N$ for (arbitrary)
given trajectories of all $K$ searchers can be written as
\begin{equation}
\label{func} \Psi\Big(\{s_k(N)\}\Big) = \prod_{k=1}^K
\prod_{n=0}^N \left[1 - I\Big(s_k(n)\Big)\right], \;\;\; I(s) =
\displaystyle \left\{\begin{array}{ll}
\displaystyle 1,  \;\;\;   s = 0, \nonumber\\
\displaystyle 0,  \;\;\;   s \neq 0.
\end{array}
\right.
\end{equation}
This indicator function has the correct behavior, namely, for this
particular set of trajectories $\Psi=0$ if any searcher hits the target
and $\Psi=1$ if none do.
All searchers are assumed to move independently. Averaging
$\Psi\Big(\{s_k(N)\}\Big)$ over all trajectories and initial
positions thus yields the probability $P_N$ that none of the searchers
has ever reached the site $s = 0$ up to time $N$,
\begin{eqnarray}
P_N &=& \left[\frac{1}{M} {\sum_{s(0)}}^\prime
E_{s(0)}\left\{\prod_{n=0}^N \left[1 - I\Big(s(n)\Big)\right]\right\}
\right]^K  \nonumber\\
&=& \left[1 - \frac{1}{M} {\sum_{s(0)}}^\prime
E_{s(0)}\left\{1 - \prod_{n=0}^N \left(1 -
I\Big(s(n)\Big)\right)\right\} \right]^K,
\end{eqnarray}
where $E_{s(0)}\left\{ \ldots \right\}$ stands for the average over
all trajectories starting at site $s(0)$, and the prime on the sum
indicates the exclusion of the target site $s(0)=0$.
Note that the summation extends over
all non-target sites of the \textit{substrate}, a one-dimensional
lattice, and does not include off-lattice contributions.
In the thermodynamic limit we set $K,M \to \infty$
keeping their ratio fixed, $K/M = \rho =$ density of searchers, to
obtain
\begin{equation}
\label{survival3}
P_N = \exp\left[- \rho {\sum_{s(0)}}^\prime
E_{s(0)}\left\{1 - \prod_{n=0}^N \left[1 -
I\Big(s(n)\Big)\right]\right\}\right].
\end{equation}
Since $s(n)$ is a homogenous process,
equation~(\ref{survival3}) can be rewritten as
\begin{equation}
\label{survival4}
P_N = \exp\left[- \rho  E_{0}\left\{\sum_{s=1}^\infty
\left(1 - \prod_{n=0}^N \left[1 - I\Big(s'(n) -
s\Big)\right)\right]\right\}\right],
\end{equation}
where now  $E_{0}\left\{ \ldots \right\}$ denotes an average over
trajectories of an auxiliary process $s'(n) =  s(n) - s(0)$ which
starts at the origin at $n = 0$. Note that the indicator function
$\left(1 - \prod_{n=0}^N \left[1 - I\Big(s'(n) - s \Big)
\right]\right)$ shows whether the site $s$ on the substrate has ever been
visited by the process $s'(n)$, $n= 0,1,2, \ldots, N$.
Consequently, the sum over all substrate sites is just the
realization-dependent number of distinct sites visited by an
individual searcher, and its average leads directly to
the crucial equation~(\ref{survival}).

For this proof to be useful in calculating $P_N$ we next need to
calculate
$S_N$.  For this purpose, note that the function in the exponent
of~(\ref{survival3}),
$\left\{1 - \prod_{n=0}^N \left[1 - I\Big(s(n)\Big)\right]\right\}$,
is the indicator function of the
event that $s(N)$ has at least once visited the origin, so that
\begin{equation}
{\sum_{s(0)}}^\prime E_{s(0)}\left\{1 - \prod_{n=0}^N
\left[1 - I\Big(s(n)\Big)\right]\right\} = {\sum_{s(0)}}^\prime
R_N(0|s(0)),
\end{equation}
where $R_N(0|s(0))$ is the probability that an arbitrary $N$-step
random process $s(N)$ has visited the origin at least once.
We can therefore write
\begin{equation}
P_N = \exp\left( -\rho {\sum_{s(0)}}^\prime R_N(0|s(0))\right).
\label{intermediate}
\end{equation}
Next we note that, by
definition, $R_N(0|s(0))$ can be formally represented as
\begin{equation}
R_N(0|s(0)) = \sum_{n=0}^N F_n(0|s(0)),
\end{equation}
where $F_n(0|s(0))$ is the probability that $s(n)$ arrived at the origin
\emph{for the first time} on
the $n$-th step, given that it started at site $s(0)$.
Applying the theory of recurrent events (see, e.g.,~\cite{hughes}),
the event that $s(n)$ is at site $0$ after $n$
steps can be decomposed into the $n$ mutually exclusive events
``$s(n)$ first arrived at site $0$ after $j$ steps, and subsequently
returned to site $0$ in $(n-j)$ steps." This allows us to write for the
distribution function $P(0|s(0);n)$ the discrete ``integral equation"
\begin{equation}
P(0|s(0);n) = \delta_{0,s(0)} \delta_{n,0} +  \sum_{j=1}^n
F_j(0|s(0)) P(0|0;n-j),
\end{equation}
which implies that the generating functions
\begin{eqnarray}
P(s|s(0);z) &=& \sum_{n=0}^{\infty} P(s|s(0);n) z^n, \nonumber\\
F(s|s(0);z) &=& \sum_{n=0}^{\infty} F_n(s|s(0)) z^n
\end{eqnarray}
at $s=0$ are simply related to each other,
\begin{equation}
F(0|s(0);z) = P(0|s(0);z) - \frac{\delta_{0,s(0)}}{P(0|0;z)}.
\end{equation}
It then follows that
\begin{equation}
R(0|s(0);z) = \sum_{n=0}^{\infty} R_N(0|s(0)) z^N =
\frac{1}{1-z} F(0|s(0);z),
\end{equation}
and
\begin{equation}
\label{imp} S(z) = \sum_{n=0}^{\infty} S_N z^N = \sum^\prime_{s(0)}
R(0|s(0);z) = \frac{1}{1 - z} \frac{\sum_{s}
P(s|0;z)}{P(0|0;z)}.
\end{equation}
Hence, in order to determine $P_N$ in equation~(\ref{survival})
via $S(z)$ we only need to determine $P(0|0;z)$ and
the normalization $\sum_{s} P(s|0;z)$.  We proceed with this
determination in the next section.

\section{Properties of the intermittent random walk}
\label{intermittent}

Since the searchers in our model move independently,
it suffices to focus on the properties of the walk of an individual searcher.
The probability $P(s|s(0);n)$ of being at site $s$ after $n$ steps
obeys the recurrence relation
\begin{eqnarray}
\label{ri}
\fl
P(s|s(0);n) &=& \frac{\alpha}{2}
\left[P(s-1|s(0);n-1) + P(s +
1|s(0);n-1)\right]  \nonumber\\
\fl
&&+ \frac{(1- \alpha)}{2} \left[P(s-L|s(0);n - T) + P(s + L|s(0);n
- T)\right]
\end{eqnarray}
for $n = 0,1,2, \ldots, N$.  Jumps between
nearest-neighbouring sites occur in one unit of time, while long-range
jumps over distance $L$ require an integer time $T$.
Equation~(\ref{ri}) therefore defines a non-Markovian process with
a memory. Since the intermittent random walks are homogeneous so that
$P(s|s(0);n) = P(s-s(0)|0;n)$, without loss of generality we henceforth
set $s(0) = 0$.

The Fourier-transformed generating function
\begin{equation}
\label{defn}
\Phi(k,z) = \sum_s \exp(i k s) P(s|0;z)
\end{equation}
can easily be calculated by multiplying both
sides of equation~(\ref{ri}) by $\exp(i k s)$ and $z^n$ and
summing over all substrate sites $s$ and time $n$. We readily find that
\begin{equation}
\label{phi} \Phi(k,z) = \left[1 - \alpha z \cos(k) - (1 -
\alpha) z^T \cos(k L)\right]^{-1}.
\end{equation}
Consequently, the lattice Green function (or site occupation
generating function) of the intermittent walk is given by
\begin{equation}
\label{2}
P(s|0;z) = \frac{1}{\pi} \int^{\pi}_{0}
\frac{\cos(k s) dk}{1 - \alpha z \cos(k) - (1 - \alpha) z^T
\cos(k L)}.
\end{equation}

One of the two quantities required for $S(z)$ in equation~(\ref{imp})
follows immediately from equations~(\ref{defn}) and (\ref{phi}) by
setting $k=0$,
\begin{equation}
\label{imp2}
\sum_{s} P(0|s;z) = \Phi(k = 0,z) = \left(1 - \alpha z - (1 -
\alpha) z^T\right)^{-1}.
\end{equation}
This leaves only the evaluation of $P(0|0;z)$ for the calculation of
$S(z)$ and consequently of $S_N$ and $P_N$ [cf.
equation~(\ref{survival})].  We shall return to this calculation
and to our goal of maximizing the detection probability of the target
presently, but first we briefly digress to obtain some associated
interesting properties of the underlying distribution $P(s|0;n)$.

\subsection{Second and fourth moments of the distribution $P(s|0;n)$}
\label{secondfourth}

Although we do not use them directly in the evaluation of $P_N$, it is
instructive to explicitly consider second and the fourth moments of the
distribution function $P(s|0;n)$ because they provide some insight into
some aspects of optimal behavior.

Consider the second moment of the distribution. One expects that
at long times the mean square displacement of a particle
will be diffusive, that is, that the second moment
$\overline{s^2(n)}$ for a particle starting from the origin
should grow proportionally with time $n$.
The overline denotes an average over many trajectories.
We differentiate $\Phi(k,z)$ of Eq.(\ref{phi}) twice with respect to
$k$ and set $k = 0$ to find that the generating function of the
second moment obeys
\begin{equation}
\overline{s^2(z)} = \sum_{n=0}^{\infty} z^n \overline{s^2(n)} =
\frac{\alpha z + (1 - \alpha) z^T L^2}{\left(1 - \alpha z - (1 -
\alpha) z^T\right)^2}.
\end{equation}
As $z \to 1^-$ the leading contribution to the generating function is of
order ${\cal O}[(1-z)^{-2}]$, which leads to linear growth in time,
$ \overline{s^2(n)} = 2 D n$, with diffusion coefficient
\begin{equation}
\label{D} D = \frac{1}{2} \frac{\alpha + (1 - \alpha)
L^2}{\left(\alpha + (1 - \alpha) T\right)^2}.
\end{equation}
Note that for $\alpha = 1$ and $\alpha = 0$ this reduces respectively to
the standard results $D = 1/2$ and $D = V^2/2$.

One must now distinguish between two situations, the cases $T = 1$
and $T > 1$. In the first case, when the flight over distance $L$
requires one unit of time precisely as does a nearest-neighbour step,
$D$ is a monotonically decreasing function of $\alpha$. It is largest
when $\alpha = 0$, when all jumps are flights over a distance $L$
requiring a unit of time, and achieves its minimal value $1/2$ when
$\alpha = 1$, when there are no long-range flights. This is, of
course, the behaviour one would expect since in the case $T = 1$
the particle is \textit{always} on the lattice. On the other
hand, when $T > 1$ and the particle spends some portion of its time
off-lattice, the situation is completely different. Here, D in
equation~(\ref{D}) shows a non-monotonic behaviour as a function of
$\alpha$.
Differentiating Eq.(\ref{D}) with respect to $\alpha$, we find that
$D$ has a maximum when
\begin{equation}
\label{opt} \alpha = \alpha^{(2)} = \frac{T + L^2 (T - 2)}{(T -
1) (L^2 - 1)},
\end{equation}
where the superscript ``2" stresses that this value of $\alpha$
only corresponds to the maximum of the second moment and not necessarily
of other moments. Note that $0 <
\alpha^{(2)} < 1$, which implies that here, contrary to the case $T
=1$, the greatest mean square displacement is
achieved by an \textit{intermittent} random walk, that is, one that
includes nearest-neighbour steps as well as off-lattice flights over a
distance $L$.
The value of $D$ corresponding to $\alpha = \alpha^{(2)}$ is given
by
\begin{equation}
D_{max} = \frac{1}{8} \frac{\Big(L^2 - 1\Big)^2}{(T - 1)
(L^2 - T)}.
\end{equation}
For large $L$, $D_{max}$ may be made very large since $D_{max} \sim
L^2/T = L V$. On the other hand, there is a penalty to pay, since
$\overline{s^2(n)} = 2 D n$ is an
asymptotic result which holds only for sufficiently long times that
increases with increasing $L$.
By increasing $L$ and optimizing $\alpha$ we can thus make
$D$ arbitrarily large, but to observe this regime we will have to
wait a progressively longer time.

Consider next the fourth moment $\overline{s^4(n)}$ of
the distribution function. Its generating function obeys
\begin{equation}
\fl
\overline{s^4(z)} = \sum_{n=0}^{\infty} z^n \overline{s^4(n)} =
\frac{6 \left(\alpha z + (1-\alpha) L^2 z^T\right)^2}{\left(1-\alpha
z - (1-\alpha) z^T\right)^3} + \frac{\alpha z + (1-\alpha) L^4
z^T}{\left(1-\alpha z - (1-\alpha) z^T\right)^2}.
\end{equation}
The leading behavior of the generating function near $z \to 1^-$
is of order ${\cal
O}[(1-z)^{-3}]$, so that at sufficiently long times
$\overline{s^4(n)}$ is quadratic in $n$,
\begin{equation}
\label{44} \overline{s^4(n)} \sim \frac{3 \left(\alpha +
(1-\alpha) L^2\right)^2}{\left(\alpha + (1-\alpha) T\right)^3}\; n^2.
\end{equation}
We point out again that the coefficient of $n^2$
shows a completely different behavior depending on
whether $T=1$ or $T > 1$. In the former case, when particles spend
all of their time on the lattice, it is a monotonic function of $\alpha$
with a maximum at $\alpha = 0$.
On the other hand, for $T > 1$ the coefficient is again a
non-monotonic function of $\alpha$.
The maximum is achieved at
\begin{equation}
\alpha^{(4)} = \frac{2 T + L^2 (T - 3)}{(T - 1) (L^2 - 1)}.
\end{equation}
For $T > 2$, the extremum $\alpha^{(4)}$ is also well
within the intermittent regime.

Note that $\alpha^{(4)} \neq
\alpha^{(2)}$, (or more precisely, $\alpha^{(4)} < \alpha^{(2)}$).
This alerts us to the fact that there is not a single ``optimal choice"
of $\alpha$.  The choice clearly depends on the property
or strategy that one wishes to optimize, and for more complex properties
it is expected to depend on the maximal search time $N$.  We are clear on
the particular strategy that we wish to optimize, namely, the
probability that the target will be detected by at least one of the
searchers by time $N$.  It is unlikely that either $\alpha^{(2)}$
or $\alpha^{(4)}$ provide this optimization.  Indeed, we find below that
the optimal detection strategy is achieved with an $N$-dependent value of
$\alpha$ that also depends on the jump velocity.

\subsection{Probability of return to the origin and expected
number of distinct sites visited}
\label{return}

We now return to the calculation at the end of
section~\ref{intermittent} to find  the generating function of the
expected value of the number of distinct sites visited by an
individual searcher. With the substitution of equation~(\ref{imp2}) into
equation~(\ref{imp}), this generating function is given by
\begin{equation}
\label{sites3}
S(z) = \frac{1}{(1-z)\left(1 - \alpha z - (1 -
\alpha) z^T\right) } \frac{1}{P(0|0;z)}.
\end{equation}
We thus turn to the analysis of $P(0|0;z)$.

We start by noting that equation~(\ref{2}) can be written as
\begin{equation}
\label{2r}
P(s|0;z)  = \int^{\infty}_{0} dt e^{- t} I_s(\zeta_1,\zeta_L),
\end{equation}
where $I_{s}(\zeta_1,\zeta_L)$ is a particular
two-variable case of a multi-variable generalized modified Bessel
function~\cite{dattoli},
\begin{equation}
I_{s}(\zeta_1,\zeta_2,\ldots) = \frac{1}{\pi} \int^{\pi}_{0}
dk \cos(k s) \exp\left[ \sum_{m=1}^{\infty} \zeta_m \cos(m k)
\right],
\end{equation}
in which $\zeta_1 = \alpha z t$, $\zeta_L = (1- \alpha) z^T t$,
and all other $\zeta_m$ are zero.  Furthermore,
$I_0(\zeta_1,\zeta_L)$ obeys
\begin{equation}
\label{L}
I_0(\zeta_1,\zeta_L) = I_0(\alpha z t) I_0((1- \alpha) z^T t)
+ 2 \sum_{l=1}^{\infty} I_{ l L}(\alpha z t) I_l((1- \alpha) z^T t).
\end{equation}
Subsitituting this latter expansion into the integral on the
right-hand-side of equation~(\ref{2r}) with $s=0$, performing the
integration over $t$, and using the integral representation
\begin{equation}
I_n(p) = \frac{(-1)^n}{\pi} \int^{1}_{-1}
\frac{dx}{\sqrt{1-x^2}}\; e^{-px}\; T_n(x),
\end{equation}
where the $T_n(x)$ are Tchebychev polynomials of the first kind, we find
that occupation generating function for the site $s=0$ reads
\begin{equation}
\label{r}
\fl
P(0|0,z) =  \frac{1}{\pi^2} \int^{1}_{-1}
\frac{dx}{\sqrt{1-x^2}} \int^{1}_{-1} \frac{dy}{\sqrt{1-y^2}}
\frac{\left[ 1 + 2 \sum_{l=1}^{\infty} (-1)^{L l + l} T_{ L l}(x)
T_l(y)\right]}{1 + \alpha z  x + (1 - \alpha) z^T  y}.
\end{equation}
Next, expanding the kernel in a series of Tchebychev polynomials,
\begin{equation}
\frac{1}{1 + \alpha z  x + (1 - \alpha) z^T y} = \frac{1}{\alpha
z \sqrt{\tau^2 - 1}} \left[1 + 2 \sum_{n=1}^{\infty} (-\xi)^n
T_n(x)\right],
\end{equation}
where
\begin{equation}
\label{tauxi}
\tau = \tau(y) = \frac{1 + (1 - \alpha) z^T y}{\alpha z} \geq 1,
\qquad \xi = \tau - \sqrt{\tau^2-1},
\end{equation}
we obtain a closed-form expression for $P(0|0,z)$ convenient for
further analysis,
\begin{equation}
\label{n}
P(0|0;z) = \frac{1}{\pi \alpha z} \int^{1}_{-1}
\frac{dy}{\sqrt{1-y^2} \sqrt{\tau^2 - 1}} \; \frac{1 - \xi^{2 L}}{1 + 2
\xi^L y + \xi^{2 L}}.
\end{equation}
Note that it is straightforward to perform the integral in
equation~(\ref{n}) in the two ``pure" limits $\alpha = 1$ and
$\alpha =0$.
In the first case we have $\tau = 1/z$, $\xi = (1 - \sqrt{1-z^2})/z$,
and
\begin{equation}
\fl
P_{\alpha=1}(0|0;z) = \frac{1}{\pi \sqrt{1 - z^2}} \int^{1}_{-1}
\frac{T_0(y) dy}{\sqrt{1-y^2}} \left[1 + 2 \sum_{l=1}^{\infty}
(-\xi^L)^l T_l(y)\right] = \frac{1}{\sqrt{1 - z^2}}.
\end{equation}
On the other hand, when $\alpha = 0$, we have $\tau = \infty$, $\xi
= 0$, and
\begin{equation}
\label{kk} P_{\alpha=0}(0|0;z) = \frac{1}{\pi} \int^{1}_{-1}
\frac{dy}{\sqrt{1-y^2}} \frac{1}{1 + z^T y} = \frac{1}{\sqrt{1 -
z^{2 T}}}.
\end{equation}
We notice that for $T=1$ the two latter expressions coincide.
One might therefore expect that $P(0|0;z)$ and the expected number
of distinct sites visited will be non-monotonic functions of $\alpha$.

In the next two sections we explicitly calculate $P(0|0;z)$ and
ultimately the desired target detection probability for intermittent
walks.

\section{Particular case: Lindenberg-Shuler intermittent random
walks with L=2 and T =1,2}
\label{particular}

To highlight the optimization procedure, consider first the simple
case of intermittent random walks with nearest-neighbour steps and
jumps over a distance $L = 2$, first studied by Lindenberg and
Shuler in~\cite{katja1} and \cite{katja2}.
In this case the integrals in equations~(\ref{2}) and (\ref{n}) can be
performed explicitly. For $L=2$ (and $T = 1$ or $T = 2$), the
generating function for the probability of being at the origin is
given by
\begin{equation}
\label{p0}
\fl
P(0|0;z)
= \frac{1}{\left[1 + (1 - \alpha) z^T\right]} \left(
\frac{\lambda_2}{\lambda_2 - \lambda_1} \frac{1}{\sqrt{1 -
\lambda_2^2}} -  \frac{\lambda_1}{\lambda_2 - \lambda_1}
\frac{1}{\sqrt{1 - \lambda_1^2}}\right),
\end{equation}
where
\begin{eqnarray}
\fl
\lambda_1 &=& - \frac{\alpha z}{2 \left[1 + (1 - \alpha) z^T\right]}
\left[1 + \left(1 + \frac{8 (1-\alpha) z^T \left[1 + (1 - \alpha)
z^T\right]}{\alpha^2 z^2}\right)^{1/2}\right]
\nonumber\\
\fl
\lambda_2 &=& - \frac{\alpha z}{2 \left[1 + (1 - \alpha) z^T\right]}
\left[1 - \left(1 + \frac{8 (1-\alpha) z^T \Big(1 + (1 - \alpha)
z^T\Big)}{\alpha^2 z^2}\right)^{1/2}\right].
\end{eqnarray}
Consequently, in this particular case the generating function of the
expected number of distinct sites visited is given by the
closed-form expression
\begin{equation}
\label{sites4}
S(z) = \left[\frac{\left(1+(1-\alpha) z^T\right)}{(1-z)
\left(1 - \alpha z - (1 - \alpha) z^T\right) }\right[\left[
\frac{\left(\lambda_2
- \lambda_1\right) \sqrt{1 - \lambda_1^2}
\sqrt{1-\lambda_2^2}}{\lambda_2 \sqrt{1 - \lambda_1^2} -  \lambda_1
\sqrt{1 - \lambda_2^2}}\right].
\end{equation}

To find the large-$n$ behavior of the distinct number of sites visited
we need to consider the limit $z \to 1^-$ of the generating function.
Recall that we expect a leading behavior of the form $S_n \sim
f(\alpha)n^{1/2}$ because ours is basically a one-dimensional random
walk, albeit intermittent. Not only do we still need to
determine $f(\alpha)$, but we also need to ascertain the importance
of subsequent terms.
A straightforward but tedious calculation leads to the expansion
\begin{equation}
\label{sites5}
\fl
S(z)= \left(\frac{2 (4 - 3 \alpha)}{\alpha +
(1-\alpha) T}\right)^{1/2} \frac{1}{\left(1 - z\right)^{3/2}} - \frac{4 (1 -
\alpha)}{\alpha^{1/2} (4 - 3 \alpha)^{1/2}} \frac{1}{\left(1 -
z\right)} + {\cal O}\left(\frac{1}{\left(1 - z\right)^{1/2}}\right).
\end{equation}
Clearly, for any given value of $z$ no matter how close to unity,
this ``blind" expansion in powers of $(1-z)$ only makes sense
if $\alpha >0$ since the second term diverges at $\alpha=0$.
This is thus only a valid expansion from which asymptotic behavior can
be extracted if $0<\alpha\leq 1$.  At $\alpha\equiv 0$ one must go
back to the original function $S(z)$, set $\alpha=0$, and then expand:
\begin{equation}
\label{sites6}
S(z) = \left(\frac{2}{T}\right)^{1/2}
\frac{1}{\left(1-z\right)^{3/2}}  + {\cal
O}\left(\frac{1}{\left(1-z\right)^{1/2}}\right).
\end{equation}
That $\alpha=0$ is a special value that can lead to discontinuities was
already noted in~\cite{katja1}.  It is the only value of the
intermittency parameter that absolutely excludes every other site from
the random walk.

Inverting $S(z)$ in equations~(\ref{sites5}) and (\ref{sites6}) at
times $n = N$ we find that for the intermittent random walk with
nearest-neighbour steps and jumps to next-nearest-neighbour
sites the leading large-$N$ behavior of $S_N$ is
\begin{equation}
\label{sites7}
S_N = f(\alpha) N^{1/2} - c(\alpha) + {\cal
O}\left(\frac{1}{N^{1/2}}\right).
\end{equation}
Reflecting the discontinuity in $\alpha$ described above, the functions
$f(\alpha)$ and $c(\alpha)$ are discontinuous at $\alpha = 0$,
\begin{equation}
\label{lab1}
f(\alpha) =\cases{\left(\frac{8 (4 - 3 \alpha)}{\pi (\alpha + (1-\alpha) T)}\right)^{1/2}&for $0 < \alpha \leq 1$,\\
(8/\pi T)^{1/2}&for $\alpha \equiv 0$,\\}
\end{equation}
and
\begin{equation}
\label{lab2}
c(\alpha) =\cases{\frac{4 (1 -
\alpha)}{\alpha^{1/2} (4 - 3
\alpha)^{1/2}}&for $0 < \alpha \leq 1$,\\
0&for $\alpha \equiv 0.$\\}
\end{equation}
Note that this ``discontinuity" should be viewed with appropriate
caution, since $S_N$ at any {\em fixed} $N$
should be a smooth function of $\alpha$.  The discontinuous forms
of equations~(\ref{lab1}) and (\ref{lab2}) arise because we are
describing an asymptotic behavior that is strictly valid only
in the limit $N \to \infty$. In fact,
the use of the $\alpha>0$ results is only appropriate as long as the
second term in $S_N$ does not dominate the first.  In any case, these
considerations require that we recognize the $\alpha=0$ discontinuity
explicitly.

\begin{figure}[!htb]
\begin{center}
\centerline{\hspace*{.2\textwidth}(a)\hspace*{.37\textwidth}(b)
\hfill}
\begin{minipage}[c]{.45\textwidth}
\includegraphics[width = 0.9\textwidth,height=0.9\textwidth]{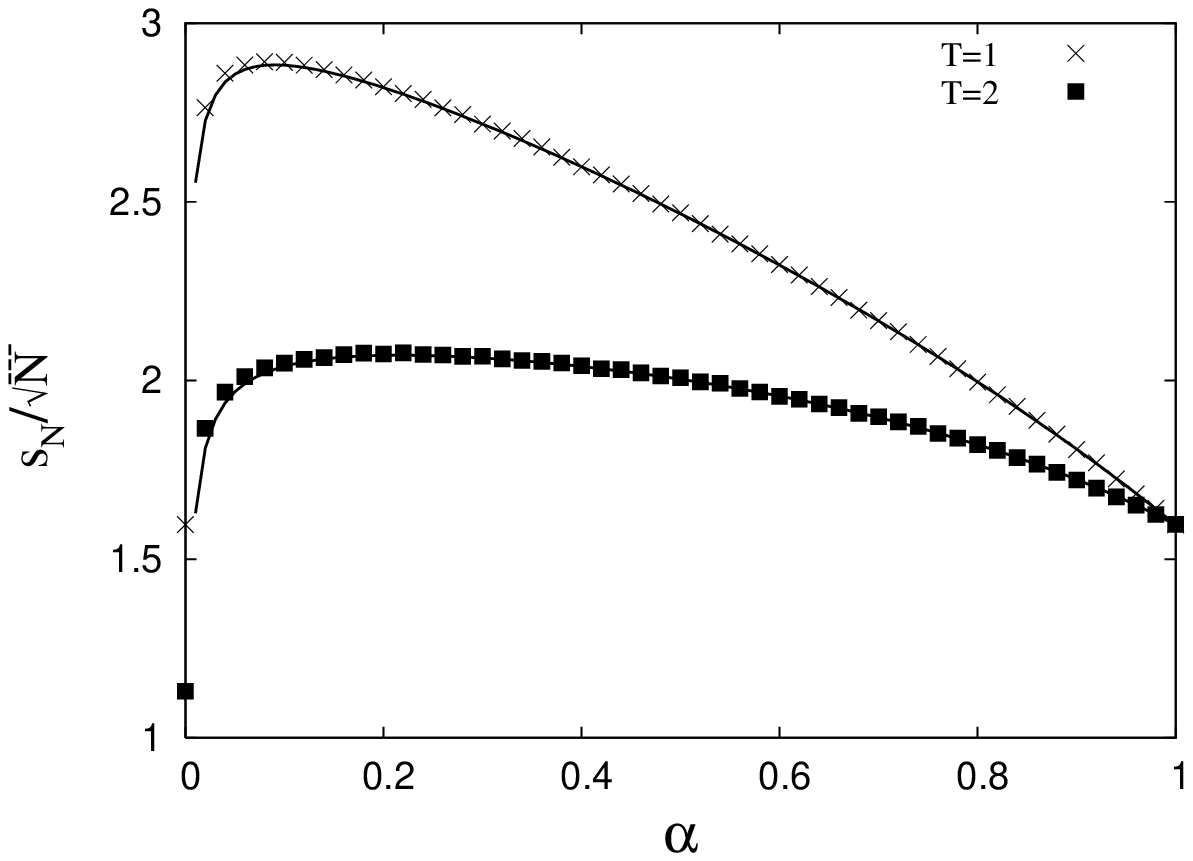}
\end{minipage}
\begin{minipage}[c]{.45\textwidth}
\includegraphics[width = 0.9\textwidth,height=0.9\textwidth]{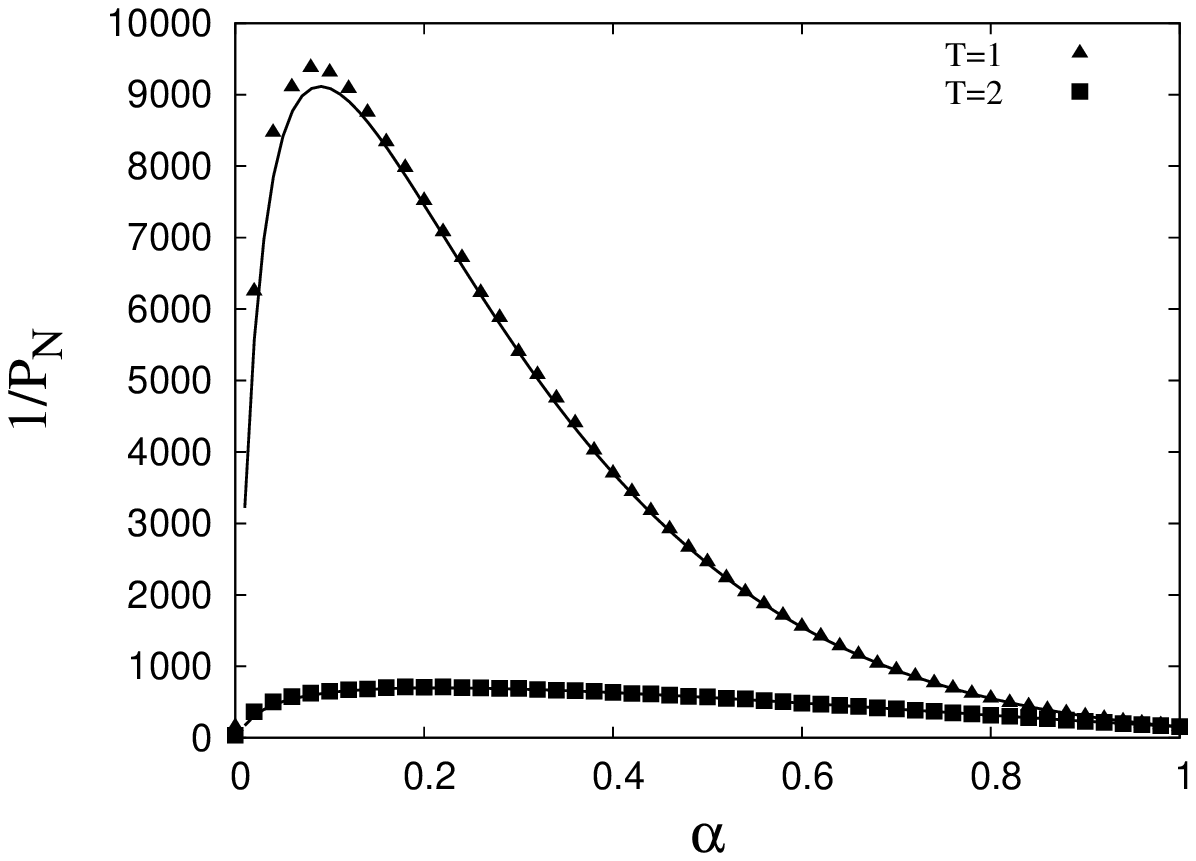}
\end{minipage}
\caption{ \label{Fig2} (a) Plot of $S_N/\sqrt{N}$ versus
$\alpha$ at $N=1000$ for $T = 1$ and $T=2$. Solid lines are
analytic results, equation~(\ref{sites7}), while symbols denote the
results of Monte Carlo simulations. (b) Plot of the inverse of the
non-detection probability, $1/P_N$, versus $\alpha$ for
$\rho =0.1$ and $N=1000$.}
\end{center}
\end{figure}

The question of interest is whether $S_N$ has a maximum
and consequently $P_N$ a minimum as a function of $\alpha$, that is,
whether it is possible to design an optimal search strategy through the
choice of the intermittency parameter.
In figure~\ref{Fig2}(a) we plot the asymptotic result for the function
$S_N$, equation~(\ref{sites7}), and the results of Monte Carlo
simulations, for $N = 10^3$ and $T = 1$ and $2$ as a function of
$\alpha$. We see that indeed $S_N$ is a non-monotonic function of
$\alpha$, and that the asymptotic form in equation~(\ref{sites7})
is in excellent agreement with the numerical data.

Our aim is to analytically determine the value of $\alpha$
which maximizes $S_N$.  Since we expect $\alpha>0$, we use the
appropriate form of the coefficients $f(\alpha)$ and $c(\alpha)$ in
equation~(\ref{sites7}).
While the coefficient $f(\alpha)$ of the leading term in
$N$ is non-monotonic, it jumps from its smallest value at $\alpha\equiv
0$ to its largest value at $\alpha=0^+$.
For such small $\alpha$ me must
retain the second term. Differentiating $S_N$ (both contributions)
with respect to $\alpha$
to find the maximum readily leads to the optimal value
\begin{equation}
\label{aa}
\alpha_{opt}(N) \approx T \left(\frac{2 \pi}{(4-T)^2
N}\right)^{1/3}
\end{equation}
with additional contributions of lower order in $N$.
Consequently, the maximal expected number of sites visited by the
intermittent random walk is attained at $\alpha = \alpha_{opt}(N)$
defined by equation~(\ref{aa}),
\begin{equation}
\label{LS}
\fl
{\rm max}_{\alpha} S_N = \left(\frac{32 N}{\pi T}\right)^{1/2} - (4 -
T)^{1/3} \left(T^{1/6} + \frac{1}{2 T^{4/3}}\right) \left(\frac{32
N}{\pi T}\right)^{1/6} + o\left(N^{1/6}\right).
\end{equation}

Let us recap the ingredients of this result and analyze its
significance. Because the optimal $\alpha$ for the most efficient
search is positive definite for any given $N$, we needed to use the
asymptotic results for $S_N$ that included the $c(\alpha)$
contribution.  The actual asymptotic number of distinct sites
visited in this optimal search, ${\rm max}_{\alpha} S_N = \left(32
N/\pi T\right)^{1/2}$, is simply the value of the first term in
equation~(\ref{sites7}) with the $\alpha>0$ form in
equation~(\ref{lab1}), when $\alpha$ is set to zero. However, the
only reason we could use this form is precisely because $\alpha$ is
not identically zero.  The actual contribution of the second term to
$S_N$ turns out to be negligible, since it contributes in
equation~(\ref{aa}) only to order $N^{1/6}$, but we need the
existence of this term to obtain the correct asymptotic result.

In this particular case of a search for an immobile hidden
target by an intermittent random walk with nearest-neighbour and
next-nearest-neighbour steps the best strategy can be summarized as
follows. If we are ready
to wait a sufficiently long time $N$ but want to be sure that at
that moment of time we will attain the lowest possible non-detection
probability $P_N$, we choose $\alpha = \alpha_{opt}(N) = T (2
\pi/(4-T)^2 N)^{1/3}$. Note also that this strategy is quite
efficient considering that nearest-neighbour and
next-nearest-neighbour steps are not very different from one
other. The enhancement of the expected number of distinct sites visited
through this strategy is a factor of $2/\sqrt{T}$.
This effect is exponentiated in the
non-detection probability, so that it is dramatically apparent in
figure~\ref{Fig2}(b).

\section{Optimal search strategy for arbitrary $L$}
\label{optimal}

The analysis presented in the previous section for a particular case
can be generalized to arbitrary values of $L$ and $T$ by following
essentially the same reasoning.  The algebraic derivations that lead to the
results are longer and more tedious, but the results are similar in form
to those obtained previously. The outcome, as we shall see, is that the
detection probability enhancement can be made much greater than that
obtained under the restrictions that $L=2$ and $T=1$ or $2$.  In fact, it
can be improved by orders of magnitude.

As before, to calculate the generating function for the distinct number
of sites visited we need to determine the generating function for the
probability of being at the origin.  The integral in equation~(\ref{n})
can no longer be done in closed form for arbitrary $L$ and $T$, but it
is possible to carry out an expansion in powers of $(1-z)$ and retain
the terms necessary for the computation of $S_N$.  As before, the cases
$0<\alpha\leq 1$ and $\alpha\equiv 0$ need to be treated separately
because the latter is still a special case.  For $\alpha\equiv 0$ the
expansion is fairly straightforward and leads to
\begin{equation}
\label{kkg}
P_{\alpha=0}(0|0;z) = \frac{1}{\sqrt{2 T}} \frac{1}{\sqrt{1-z}} + {\cal
O}\left(\sqrt{1-z}\right).
\end{equation}
The calculation is more elaborate for $0<\alpha\leq 1$.
It is essential to retain not only the leading
divergent contribution in $(1-z)$ as $z\to 1^-$, but also the contribution
independent of $(1-z)$.  This is because ultimately once again we need the
contributions of ${\cal O}(N^{1/2})$ as well as
${\cal O}(N^0)$ to $S_N$ to arrive at the correct optimization. We find
\begin{equation}
\fl \label{lul} P(0|0;z) = \frac{1}{\sqrt{2 \left(\alpha + (1 -
\alpha) T\right) \left(\alpha + (1- \alpha) L^2\right)}}
\frac{1}{\sqrt{1-z}} + I(\alpha, L) + {\cal
O}\left(\sqrt{1-z}\right),
\end{equation}
where
\begin{equation}
\fl \label{int} I(\alpha, L) = \frac{1}{\pi \alpha} \int^{1}_{-1}
\frac{dy}{\sqrt{1-y^2} \sqrt{\tau_1^2 - 1}} \left(\frac{1 - \xi_1^{2
L}}{1 + 2 \xi_1^L y + \xi_1^{2 L}} - \frac{L \sqrt{\tau_1^2 -
1}}{L^2 (\tau_1 - 1) + (y + 1)}\right),
\end{equation}
and $\tau_1$ and $\xi_1$ are the values of the
functions $\tau$ and $\xi$ of equation~(\ref{tauxi}) at $z \equiv 1$.
Substitution into equation~(\ref{sites3}) leads to
\begin{eqnarray}
\label{kl}
S(z) &=& \left(\frac{2 \left(\alpha + (1-\alpha)
L^2\right)}{\left(\alpha + (1-\alpha) T\right)}\right)^{1/2}
\frac{1}{\left(1-z\right)^{3/2}} - \frac{2 I(\alpha, L)
\left(\alpha + (1-\alpha) L^2\right)}{(1-z)} \nonumber \\
&&+ {\cal O}\left(\frac{1}{\left(1-z\right)^{1/2}}\right),
\end{eqnarray}
which implies that at $n = N \gg 1$ the leading
asymptotic behavior of the expected number of distinct sites
visited by an intermittent random walk is
\begin{equation}
\label{M} \fl S_N =\left(\frac{\left(\alpha + (1-\alpha)
L^2\right)}{\left(\alpha + (1-\alpha) T\right)}\right)^{1/2}
\left(\frac{8 N}{\pi}\right)^{1/2} - 2 I(\alpha, L) \left(\alpha +
(1-\alpha) L^2\right) + {\cal O}\left(\frac{1}{N^{1/2}}\right).
\end{equation}
Finally, the integral $I(\alpha,L)$ is not available in closed
form, but its important contribution to the problem can be estimated.
For sufficiently large $N$ it is small
everywhere except for the neighbourhood of $\alpha=0$.  To extract the
leading contribution for small $\alpha$ we change the variable of
integration in equation~(\ref{int}) from $y$ to the auxilliary variable
$u= (1 + (1- \alpha) y)/\alpha$, in terms of which
\begin{eqnarray}
\label{int1} \fl I(\alpha, L) &=& \frac{1-\alpha}{\pi \sqrt{\alpha}}
\int^{(2 - \alpha)/\alpha}_{1}
\frac{du}{(u - 1) \sqrt{(u +1) (2 - \alpha (u+1))}} \nonumber\\
\fl &&\times\left(\frac{1 - \xi_1^{2 L}}{(1-\alpha)(1 + \xi_1^{2 L})
+ 2 (\alpha u -1) \xi_1^L} - \frac{L \sqrt{u +
1}}{\sqrt{u - 1} (\alpha + (1 - \alpha) L^2}\right).
\end{eqnarray}
The singularity at the lower limit of integration $u=1$ is integrable
for any value of $\alpha$. Furthermore, the integrand vanishes sufficiently
fast to insure that it is integrable as $u \to \infty$. Hence, we may set
$\alpha \equiv 0$ everywhere in the integrand and we can extend the
upper integration limit to infinity. This provides the leading
small-$\alpha$ contribution,
\begin{equation}
\label{int2} \fl I(\alpha, L) \sim \frac{(1 - \alpha) g(L)}{\pi
\sqrt{\alpha}}, \quad g(L) = \int^{\infty}_0 \frac{d
\phi}{\sinh(\phi)} \left(\coth(L \phi) - \frac{1}{L}
\coth(\phi)\right).
\end{equation}

The subsequent reasoning now proceeds exactly as in the previous section.
We wish to
determine the value of $\alpha$ that maximizes $S_N$. Again, we expect
that this $\alpha$ is small but strictly positive, so we focus on
the result~(\ref{M}). Maximizing $S_N$ with respect to $\alpha$ we find
that
\begin{equation}
\label{opti}
\alpha_{opt}(N) \simeq \frac{L^{7/3}
g^{2/3}(L)}{V^{1/3} (L V - 1)^{2/3}} \frac{1}{(8 \pi N)^{1/3}},
\end{equation}
The resulting asymptotic form for the distinct
number of sites visited finally is
\begin{equation}
\label{tr}
{\rm max}_{\alpha} S_N = \left(L V\right)^{1/2}
\left(\frac{8 N}{\pi}\right)^{1/2} - \gamma(L) \left(\frac{8
N}{\pi}\right)^{1/6}   +  o\left(N^{1/6}\right),
\end{equation}
where
\begin{equation}
\gamma(L) = \frac{(4 \pi + 1)}{2 \pi^{2/3}} \Big(V g^4(L) L^5 (L
V - 1)^2 \Big)^{1/6}.
\end{equation}
This is the generalization of equation~(\ref{LS}).  It is valid when
$\alpha_{opt}(N) \ll 1$, that is, when $N\gg g^2(L)L^2$, which insures
that the first term in ${\rm max}_{\alpha} S_N$ is dominant.
Note that the leading contribution to $S_N$
is again the outcome of setting $\alpha=0$ in the first term
of~(\ref{M}).  The second term ultimately does not contribute to leading
order in $N$, but, to stress once again,
we were required the use the $\alpha>0$ result~(\ref{M})
because $\alpha_{opt}(N)$ is positive definite.

\begin{figure}[!htb]
\begin{center}
\centerline{\hspace*{.2\textwidth}(a)\hspace*{.37\textwidth}(b)
\hfill}
\begin{minipage}[c]{.45\textwidth}
\includegraphics[width = 0.9\textwidth,height=0.9\textwidth]{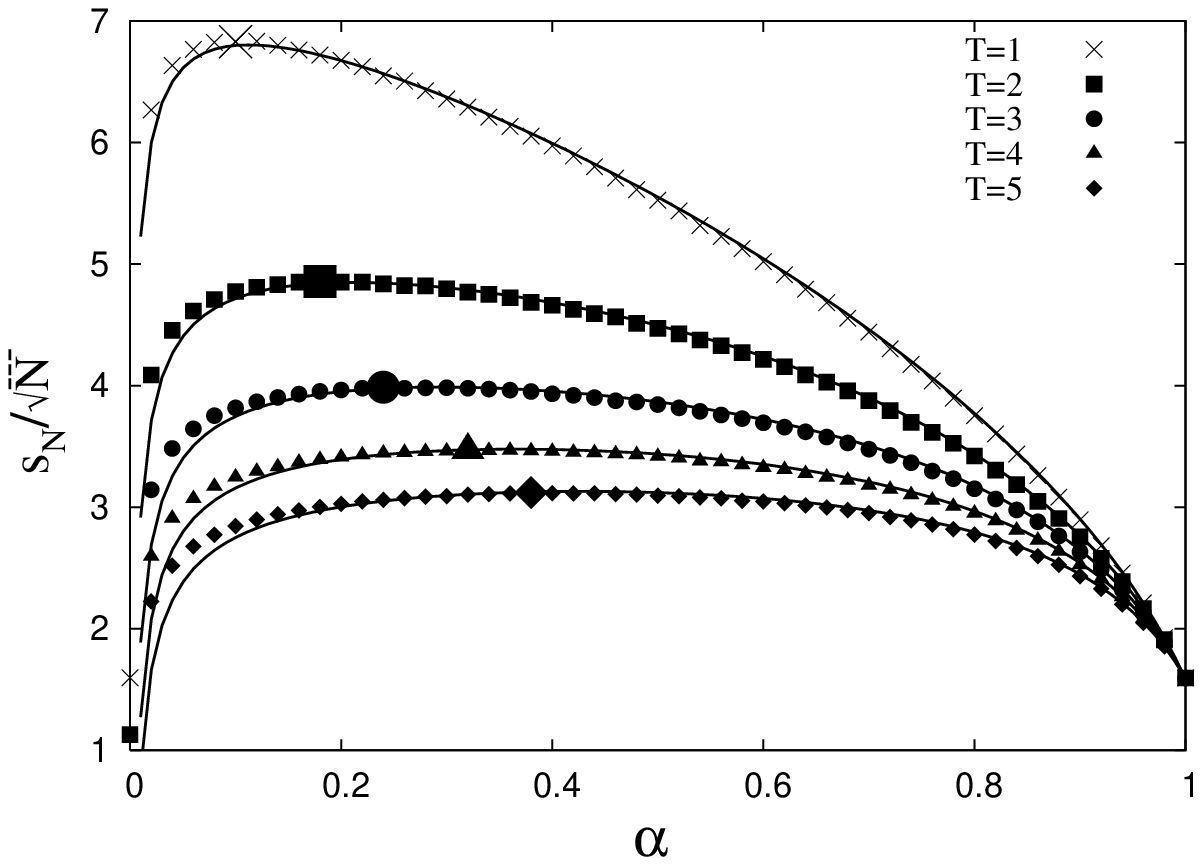}
\end{minipage}
\begin{minipage}[c]{.45\textwidth}
\includegraphics[width = 0.9\textwidth,height=0.9\textwidth]{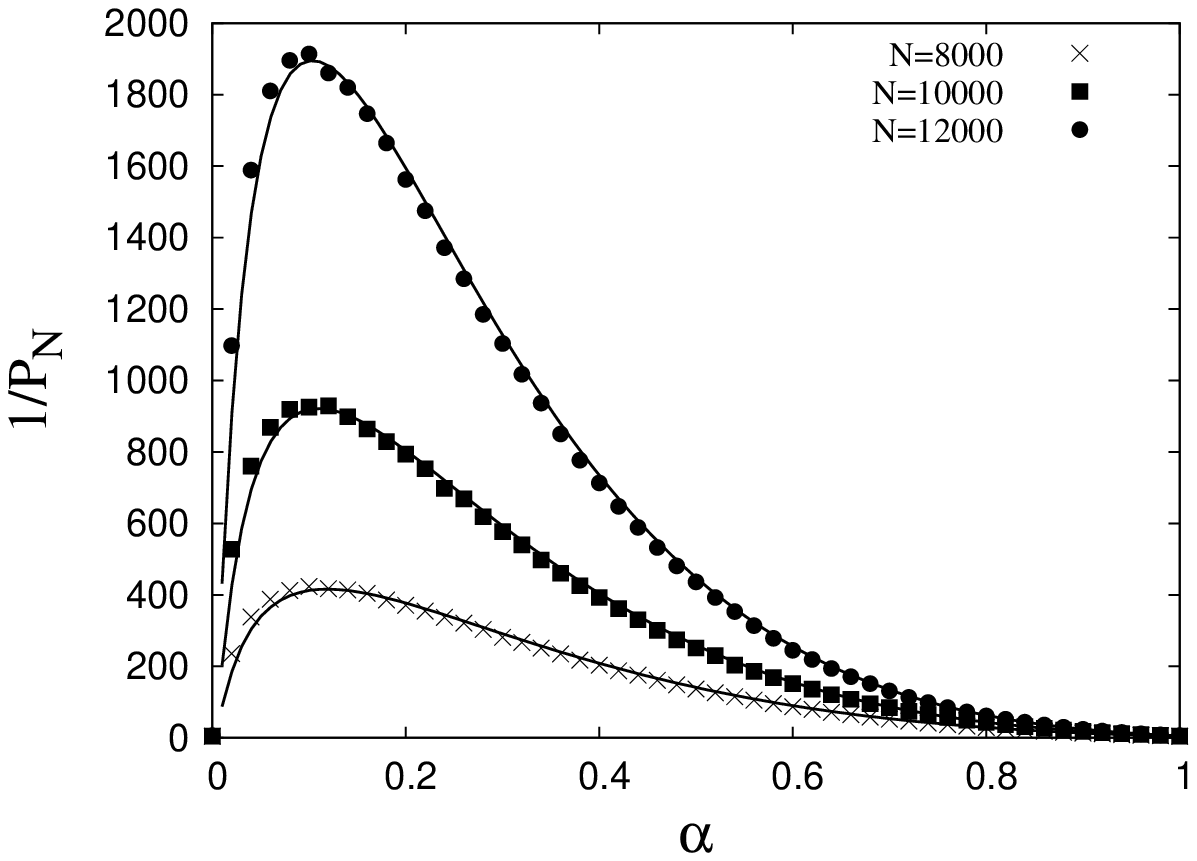}
\end{minipage}
\caption{ \label{Fig3} (a) Plot of $S_N/\sqrt{N}$ versus
$\alpha$ for $L = 5$ at $N=10000$, for $T = 1,2,3,4,5$. Solid lines are
analytic results, equation~(\ref{M}) with (\ref{int2}), while
symbols denote the results of Monte Carlo simulations. The larger
symbols denote the positions of the maxima. (b) Plot of the inverse
of the non-detection probability, $1/P_N$, versus $\alpha$ for
$\rho =0.01$ and $N=8000, 10000$, and $12000$.}
\end{center}
\end{figure}

In figure~\ref{Fig3}(a) we plot our analytic result~(\ref{M})
using the estimate~(\ref{int2}) (lines) as well as the results of
Monte Carlo simulations (symbols) as a function of $\alpha$.  The
results are shown for $L=5$ and various values of $N$ and $T$, and the
agreement is again excellent.  The larger symbols in the figure denote
the positions of the maxima, which are well captured by our analytical
curves.  Figure~\ref{Fig3}(b) shows the inverse of
the associated non-detection probability for various values of $N$.

The essential result of this section is the very large enhancement
of the target detection probability that is possible by the appropriate
$N$-dependent choice of the intermittency parameter compared to the
outcome for the
``pure" random walks with $\alpha = 0$ or $\alpha = 1$.
The leading term in equation~(\ref{tr}) contains the factor
$\sqrt{L V}$ which may be large and may therefore lead to a
substantial decrease in the probability that the target has not been
detected  up to time $N$. For example, for $L =5$, $T=1$, and $N =
10000$, with a density of searchers as low as $\rho=0.01$
(cf. figure~\ref{Fig3}(b)), the non-detection
probability for the ``pure" cases is $P_N \approx 0.2$. On
the other hand, if we choose, with $\alpha = \alpha_{opt}(10000)
\approx 0.07$,
the probability that the target has not been yet detected up to this
time is $P_N \approx 0.0003$, i.e., three orders of magnitude smaller.
Note finally that in one dimension for fixed $L$
the optimal search strategy involves a progressively smaller fraction of
nearest neighbour steps as $N$ is increased.

\section{Conclusions}
\label{conclusions}

We have studied the search kinetics of a single target
hidden on a one-dimensional regular lattice by a concentration of
``searchers" performing \textit{intermittent} random walks; that is,
at each tick of the clock, each searcher has a choice of stepping on a
nearest neighbour with probability $\alpha$, or flying off-lattice in
either direction with velocity $V$ over a distance $L$
with probability $(1-\alpha)$.  We have
determined the probability $P_N$ that the target remains
\textit{undetected} up to the maximal search time $N$, have
established that $P_N$ is a non-monotonic function of $\alpha$,
and that in fact it has a sharp minimum at an $N$-dependent
value of $\alpha$.
Consequently, we have shown that the search efficiency can be
dramatically enhanced, by orders of magnitude, by choosing
$\alpha$ appropriately.

We close this paper with some observations and problems for future
research.  If we are at liberty to choose
$L$, we may increase the number of distinct sites visited
within a given time interval $N$ by picking an $N$-dependent value of
$L$, cf. equation~(\ref{tr}). On
the other hand, equation~(\ref{tr}) is a parabolic function of $L$, which
implies that there should be an optimal $L$ for any fixed $N$. To wit,
making $L$ too large for fixed $N$ would be counter-productive
since the searchers would spend most of their time flying
off-lattice.  On the other hand, a value of $L$ that is too small would
detain the searchers in already well-explored regions and delay the
exploration of new ones. This issue requires a more subtle
analysis and a more intricate optimization procedure, to be
discussed elsewhere \cite{we2}.

\begin{figure}[!htb]
\begin{center}
\includegraphics[width = 0.6\textwidth,height=0.4\textwidth]{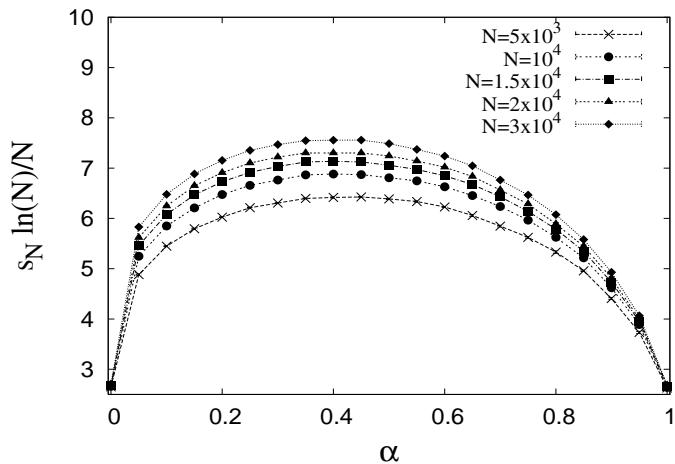}
\caption{ \label{Fig4} Results of Monte Carlo simulations on a
two-dimensional square lattice. Plot of $S_N ln(N)/N$ versus
$\alpha$ for different $N$. Symbols denote the results of the
simulations, and the solid lines are just guides for the eye.}
\end{center}
\end{figure}

Our model can be straightforwardly extended to higher dimensions. In
two dimensions there is also a value of $\alpha$ that maximizes
$S_N$ and that depends on the maximal search time $N$, as seen in
the numerical results shown in figure~\ref{Fig4}.  In three
dimensions and higher, we expect an $N$-independent value of
$\alpha$ to maximize $S_N$.

The results in this paper were obtained for the case of
perfect detection ($q=1$), that is, once a searcher lands at the site
occupied by the target it detects it with certainty. It is
easy to show that in the general case when detection of the target
upon encouter is less than certain ($0 \leq q < 1$),
the probability that the target remains undetected up to time $N$ is
\begin{equation}
P_N = \exp\left(-\rho Q_N\right),
\end{equation}
where $Q_N$ is determined via its generating function,
\begin{equation}
Q(z) = \sum_{N=0}^{\infty} Q_N z^N =  \frac{1}{1-z} \frac{q}{\left[q
P(0|0;z) + 1 - q\right]} \sum_s P(s|0;z).
\end{equation}
In low dimensional systems, i.e., in one and
two dimensions, $P(0|0;z)$ diverges as $z \to 1^{-}$, and
consequently, in the limit $N \to \infty$, $Q_N$ converges to
$S_N$ so that the leading large-$N$ behavior will be independent
of $q$ provided that $q > 0$. In three and higher dimensions
the leading behavior will depend on $q$.

An interesting extension of our model may be a situation in which
the target itself moves randomly. It is known, however,
that if this motion is diffusive (or sub-diffusive) in low
dimensions, the long-time asymptotic form of the probability $P_N$
is exactly the same as when the target is immobile
(see~\cite{bray} and also~\cite{osh,santos}). Consequently, in
low dimensions in the large-$N$ limit, the search for a diffusive (or
sub-diffusive) target by searchers performing intermittent random
walks will proceed in exactly the same way as determined in this
paper for the case of an immobile target.

Finally, we remark that a sensible extension of this model would be
to introduce the possibility of jumping, with a given velocity over
arbitrary distances $l=1,2,3, \ldots$, with probability $p_l$. A
robust approach would be to search for the distribution $p_l$
that minimizes the non-detection probability $P_N$.

\section{Acknowledgements}

The authors wish to thank D.Piorunsky, O.B\'enichou, M.Moreau and
R.Voituriez for stimulating discussions and useful comments, as well
as O.Vasilyev for the help with numerical simulations. K.L. is supported
in part by the National Science Foundation under grant PHY-0354937.

\pagebreak

\end{document}